\documentclass[11pt,a4paper]{article}
\usepackage[hyperref]{myacl}
\usepackage{times}
\usepackage{latexsym}

\usepackage{url}

\usepackage[utf8]{inputenc}
\usepackage{epigraph}
\usepackage{graphicx}
\usepackage{amsmath}
\usepackage{booktabs}
\usepackage{bbding}
\usepackage{colortbl}
\usepackage{float}
\usepackage{makecell}
\usepackage{relsize}

\definecolor{White}{rgb}{1,1,1}
\definecolor{HighlightGreen}{rgb}{0.58,0.71,0.29}
\definecolor{HighlightBlue}{rgb}{0.35,0.4,0.51}

\aclfinalcopy % Uncomment this line for the final submission

\restylefloat{figure}
\title{The DALPHI annotation framework \& how its pre-annotations can improve annotator efficiency}

\author{Robert Greinacher$^{1,2}$ \& Franziska Horn$^3$\\
  $^1$Cognitive Psychology and Cognitive Ergonomics, Technische Universität Berlin, Berlin, Germany\\
  $^2$Institute of Neuroscience \& Psychology, University of Glasgow, Glasgow, UK\\
  $^3$Machine Learning Group, Technische Universität Berlin, Berlin, Germany\\
  {\tt research@robert-greinacher.de}}

\date{}

\begin{document}
\maketitle
\begin{abstract}
Producing the required amounts of training data for machine learning and NLP tasks often involves human annotators doing very repetitive and monotonous work. In this paper, we present and evaluate our novel annotation framework DALPHI, which facilitates the annotation process by providing the annotator with suggestions generated by an automated, active-learning based assistance system. In a study with 66 participants, we demonstrate on the exemplary task of annotating named entities in text documents that with this assistance system the annotation processes can be improved with respect to the quality and quantity of produced annotations, even if the pre-annotations provided by the assistance system are at a recall level of only 50\%.
\end{abstract}

\section{Introduction}
Machine learning (ML) applications often require huge amounts of training data, especially when deep neural networks are involved. While for natural language processing (NLP) tasks such as named entity recognition (NER), large corpora of annotated text documents are usually available for the English language, other languages or specific domains are lacking the required resources to be tackled with advanced ML algorithms.

Generating large quantities of training data, e.g.,\ by annotating named entities in text documents, is essential, but at the same time also a very monotonous and time consuming task.
We therefore challenge how this annotation process is implemented in currently available annotation tools and developed a new general purpose and open source annotation framework called DALPHI to accelerate and improve the annotation of documents. Our system is customizable to fit any kind of data and desired user interface (UI). Most importantly, it provides an annotation work cycle where the human annotators are supported by a constantly improving active-learning (AL) based assistance system, which supplies pre-annotations using a task specific ML unit.

Our main hypothesis was that overseeing suggestions and correcting them as appropriate compared to manually annotating a complete document will lower the workload and therefore improve and accelerate the annotation task. However, as these pre-annotations are generated by an ML model trained on previously annotated data, they are most likely far from perfect. Therefore, we conducted a study with 66 participants to evaluate our annotation system on the exemplary task of NER, specifically to determine the level of recall our automated assistance system would have to achieve for the pre-annotations to be helpful.

\subsection{Related work}
\citet{biemann2017collaborative} provide a comprehensive list of web-based, collaborative annotation tools, most of which focus on text annotation (highlighting spans and relations between spans). Two frequently used tools are \textit{GATE Teamware}~\cite{bontcheva2013gate} and \textit{WebAnno}~\cite{yimam2013webanno,yimam2014automatic}, which already support the annotator by providing pre-annotations. However, these tools only provide limited support for custom ML models to generate pre-annotations with or for active learning setups, i.e., to automatically prioritize those documents for annotation that would be most informative for the ML model. Additionally, their UIs can not be easily customized to fit novel annotation tasks.

Previous studies already showed that annotator productivity can be enhanced by an assistance system generating pre-annotations, but most of these studies were conducted using only rule-based assistance systems \cite{day1997mixed,neveol2011semi,lingren2013evaluating} or ML models operating at uncontrolled levels of accuracy \cite{komiya2016comparison,ganchev2007semi}.
However, as ML models vary in their accuracy depending on the available training data and task, it is important to determine what level of performance such an assistance system needs to achieve for the generated pre-annotations to accelerate and improve the annotation process. While this was previously investigated for the task of part-of-speech (POS) tagging \cite{ringger2008assessing,fort2010influence}, these studies were not performed using an annotation tool but rather a synthetic testbed. Therefore it remains unclear how these results translate to real annotation settings.

\section{The DALPHI annotation framework}
\begin{figure}
  \includegraphics[width=\linewidth]{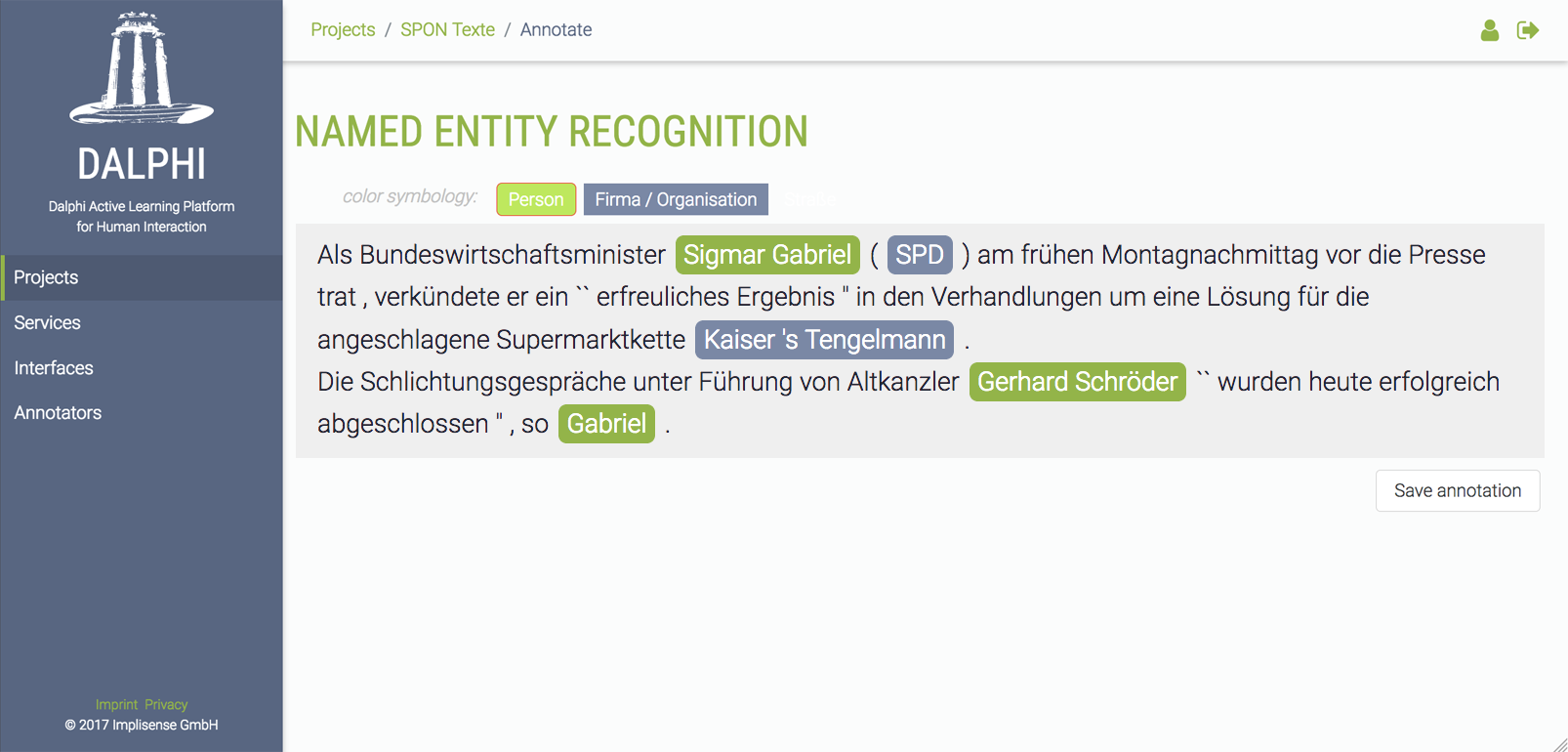}
  \caption{The DALPHI UI for text annotations.}
  \label{fig:DalphiAnnotationUIWithAnnotations}
\end{figure}
Our novel open source web-based annotation framework is called \emph{DALPHI Active Learning Platform for Human Interaction}. It can be customized to fit different data formats and to provide task specific UIs. An example setup for the annotation of text documents with named entities is shown in Fig.~\ref{fig:DalphiAnnotationUIWithAnnotations}.
\begin{figure}
  \includegraphics[width=\linewidth]{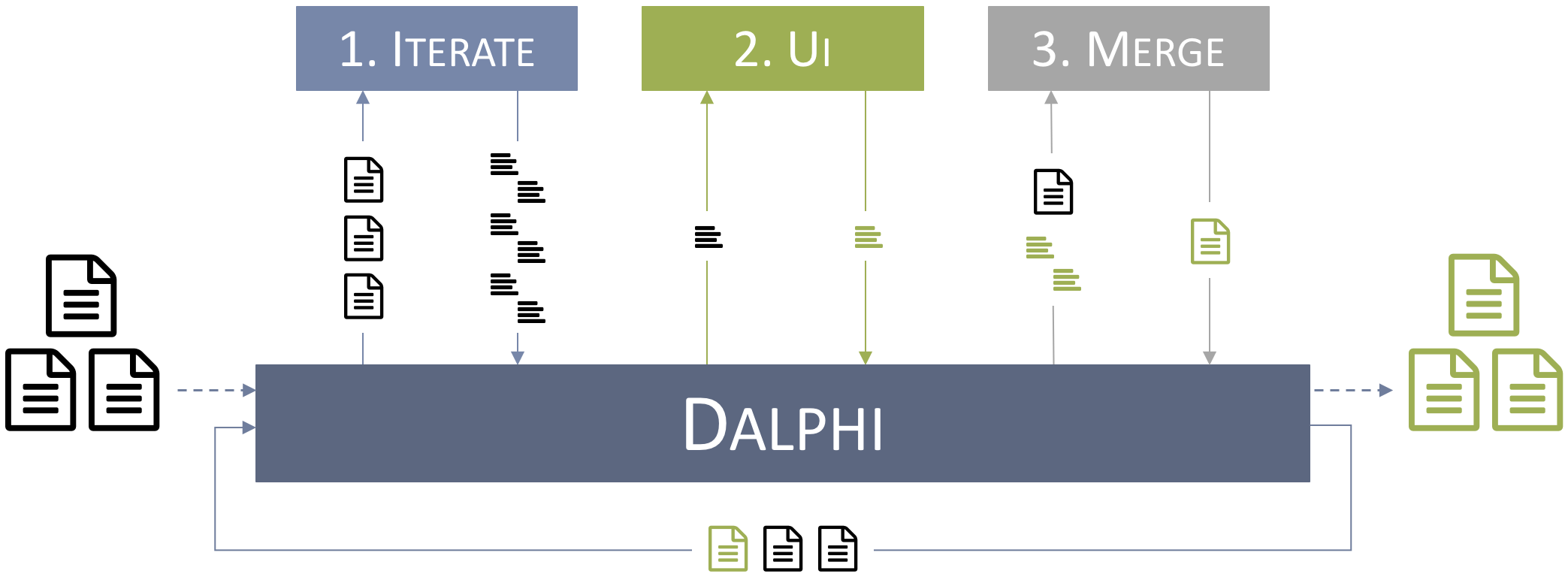}
  \caption{The processing and data flow cycle of the DALPHI framework. Light document symbols are annotated, dark are not.}
  \label{fig:DalphiWorkflow}
\end{figure}
The annotation workflow of DALPHI is structured as follows (Fig.~\ref{fig:DalphiWorkflow}): The annotation of a whole corpus is split into several iterations. First, the documents that should be annotated in the current annotation iteration are retrieved and then presented one by one, together with the corresponding pre-annotations, to the annotator in the UI. The human annotator then annotates the documents by correcting the pre-annotations and completing misses. When all documents of the current iteration are annotated, they are merged back with the rest of the corpus. Before the next iteration, the task specific ML unit can be retrained on these newly annotated documents to generate better pre-annotations for the remaining unannotated documents. Furthermore, to ensure that maximally informative training samples are annotated, the batch of documents for the next iteration can be selected using active learning, i.e., based on how certain the ML model was when generating the pre-annotations.

While in the present study we focus on generating annotations for a NER task, DALPHI can easily be used to generate training data for other tasks as well. For this, specific components of the system, such as the UI, need to be adapted, which should be straightforward given the open source collection of examples we provide together with the DALPHI implementation.
DALPHI, as well as an example setup for NER with pre-annotations generated by a classifier from NLTK~\cite{nltk2017chunkne}, is available on GitHub.\footnote{\url{https://github.com/Dalphi}}

\section{Evaluation \& Results}
An automated assistance system can not be expected to provide perfect pre-annotations.
To systematically compare the impact of pre-annotations with different levels of correctness (in terms of recall) on the human annotators, we used three assistance systems that produced correct pre-annotations 10\%, 50\%, or 90\% of the time.
%If the assistance made only 10\% correct suggestions, at the same time the remaining 90\% were erroneously annotated.
The 10\% level represents the lower bound of what we assumed might be a useful assistance and could, for example, correspond to the performance achieved by an ML model at the beginning of the annotation process when almost no training data is available. 50\% recall can realistically be achieved with an ML model trained on moderate amounts of data,\footnote{50\% recall is less than the median of the performance in the GermEval competition; the top German NER systems perform below 80\% recall~\cite{benikova2014germeval}.} which makes this scenario the most relevant in our comparison. The last system (90\%) is currently out of reach for many languages and domains. Nevertheless, studying its impact on the perceived monotony and workload is important for designing a novel, user-centred system.

To get pre-annotations at these controlled recall levels, we did not use an actual ML model, but instead simulated the assistance system by using a predefined lookup table with randomized chunk / label combinations. Nevertheless, to get realistic performances, it is very important that the \emph{types of errors} that are made by the simulated assistance system at a specific recall level match those that would be made by a real ML model. We therefore used a state of the art Conditional Random Field (CRF) NER model~\cite{lample2016neural} to predict the labels for the corpus we used in our study. We identified five possible types of annotation errors, ranging from not annotated at all to wrongly annotated in terms of label and span (Table~\ref{tab:annotationErrors}). The same distribution of errors made by the CRF classifier (\textit{Dist.}~column in Table~\ref{tab:annotationErrors}) was then used to generate the pre-annotations of the simulated assistance system.
\begin{table*}
  \centering
  \caption{Possible annotation errors. Green labels are person names, blue labels are company names.}
  \begin{tabular}{ccllc}
    \toprule
    Label & Span & Description & Example & Dist. \\
    \midrule
    \Checkmark & \Checkmark & correct & CEO \colorbox{HighlightGreen}{\textcolor{White}{Lorene Duck}} raises wages. & 57.8\% \\
    \Checkmark & \XSolidBrush & correct label, wrong span & CEO Lorene \colorbox{HighlightGreen}{\textcolor{White}{Duck raises}} wages. & 7.8\% \\
    \XSolidBrush & \Checkmark & wrong label, correct span & CEO \colorbox{HighlightBlue}{\textcolor{White}{Lorene Duck}} raises wages. & 3.1\% \\
    \XSolidBrush & \XSolidBrush & wrong label and span & CEO Lorene \colorbox{HighlightBlue}{\textcolor{White}{Duck}} raises wages. & 0.0\% \\
    \XSolidBrush & \XSolidBrush & unnecessary annotation & CEO Lorene Duck raises \colorbox{HighlightGreen}{\textcolor{White}{wages}}. & 1.5\% \\
    - & - & missing annotation & CEO Lorene Duck raises wages. & 29.8\% \\
    \bottomrule
  \end{tabular}
  \label{tab:annotationErrors}
\end{table*}

For our study we chose 14 recent German news articles (approx.\ 6000 words in total), selected by length and count of possible annotations. They were presented one by one and split into their 73 paragraphs. We compiled the collection to be as diverse as possible in order to mute performance differences of the participants with different expertise and affinities.
\begin{figure}
  \includegraphics[width=\linewidth]{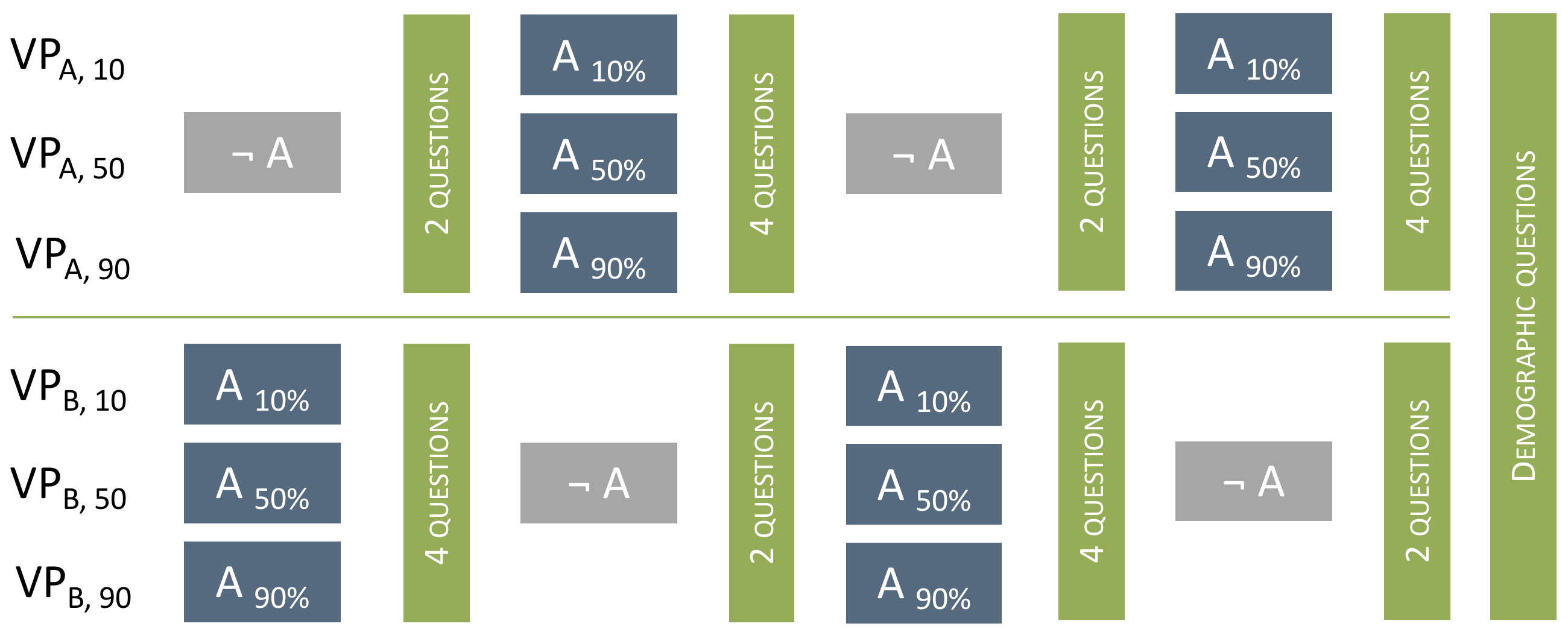}
  \caption{The phases of the study for all groups.}
  \label{fig:DesignStructure}
\end{figure}
The study was conducted in lab conditions on site with 66 participants who had no prior annotation experience.
After a short training phase, where the subjects could get familiar with the task and the UI, they then annotated the texts on their own in the main part of the experiment. This main part was divided into four blocks, where the assistance was alternating present in two out of four blocks (Fig.~\ref{fig:DesignStructure}). According to our gold standard there were 310 named entities (persons or organizations) to annotate in the entire texts.\footnote{We created this gold standard manually ourselves and it, together with other data, is available online at \url{https://gitlab.com/pasubda/dalphi-study-data/}.} Annotations were distributed as equally as possible between the four blocks. After each block, the participants were asked to fill in a questionnaire about their perceived workload and monotony of the task.
A post-processing pipeline matched the 66 differently annotated corpora with the gold standard to quantify the errors. For the statistical analysis, differences between blocks with and without assistance were calculated to assess changes using one sample t-tests.

\begin{figure}
  \includegraphics[width=\linewidth]{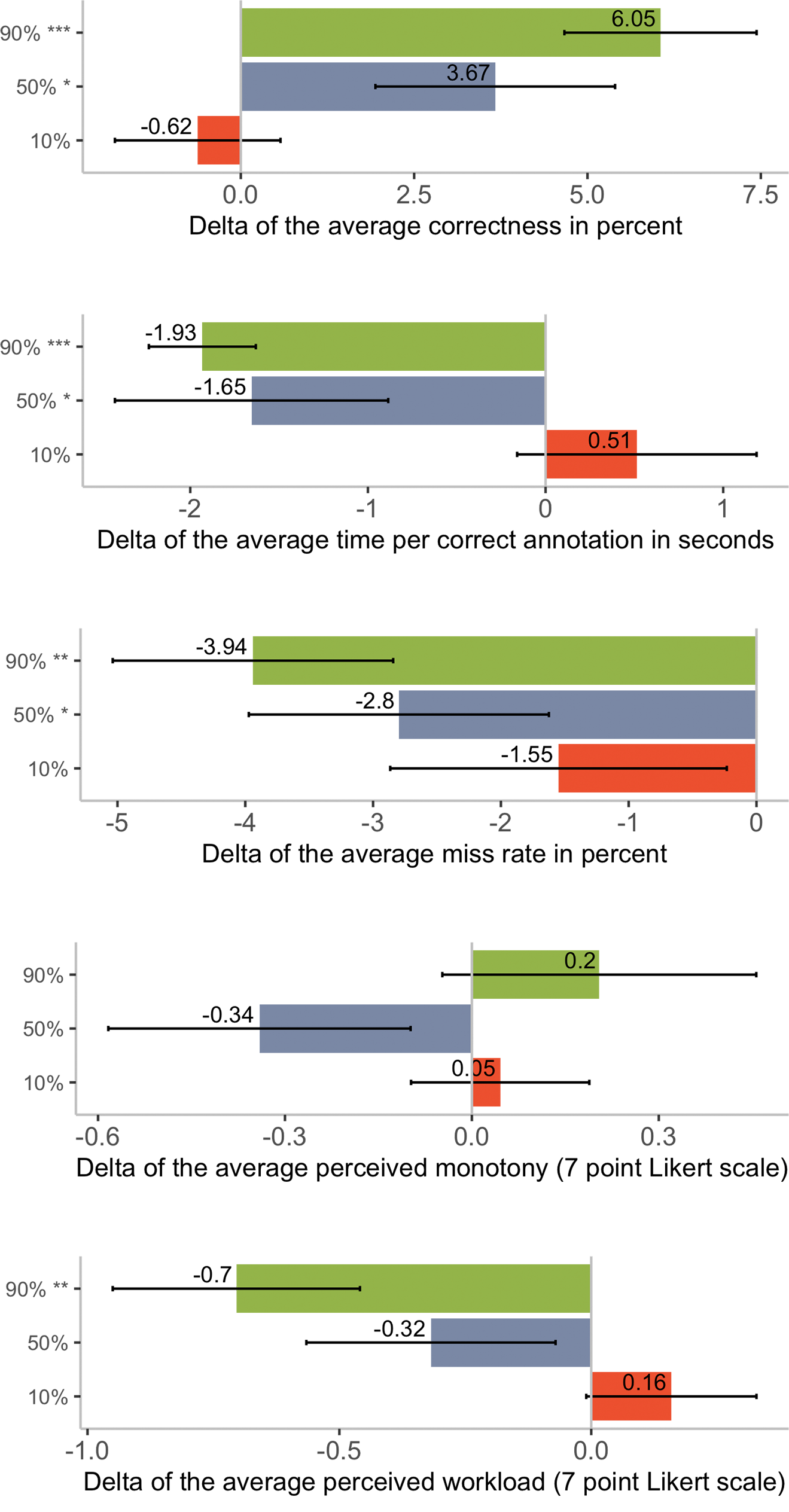}
  \caption{All three levels of the assistance (90\%: green, 50\%: blue, 10\%: red) regarding the three performance dimensions and the exploratory questionnaire data analysis. The length of each bar is the mean value of the condition, the whiskers depict the standard error of the mean in both directions.}
  \label{fig:Analysis}
\end{figure}

As shown in Fig.~\ref{fig:Analysis}, our lower-bound assistance system with 10\% correct pre-annotations did not significantly facilitate the process in any of the three performance dimensions. However, the level 50\% and level 90\% assistance systems yielded significant improvements compared to the baseline (no assistance system). Regarding the perceived workload of the task, we found a significant difference between the assistance with 90\% recall and the baseline, but none of the assistance systems significantly improved the perception of monotony of the annotation task.
\section{Discussion}
In this paper we introduced our novel annotation framework DALPHI and examined if and how the task of manual data labelling can be improved by using pre-annotations as generated by our system. In particular, we showed that it is possible to increase the performance of the annotators in terms of producing more correct and complete annotations in less time, provided that the assistance system achieves a certain level of recall. Furthermore, we used questionnaires in order to answer if the task itself can be improved: While a very advanced assistance system can reduce the perceived workload, the perceived monotony of the task remains unchanged.

An assistance system achieving 90\% recall is not realistic for many tasks. However, crafting a model that produces correct annotations in 50\% of the cases is feasible and already improves the annotation process by increasing the quality of the produced annotations (yielding an average of 87.6\% correct annotations compared to the baseline performance of 83.9\%) and reducing the number of missed annotations by 36\%.\footnote{This data describes the average performance of a single annotator. To generate labelled datasets, the annotations of several annotators are usually combined to yield better results \cite{brants2000inter}.} Furthermore, an assistance system can save time, i.e.\ money: A typical corpus for an NLP task contains tens of thousands of sentences and annotations (e.g.\ GermEval 2014: 31,297 sentences and 37,926 named entity annotations \cite{germEval2014ner}). Without any assistance, our participants needed an average of 8.2$\pm$2.3 seconds per correct named entity annotation. Annotating the GermEval corpus at this pace would have taken more than two weeks of full time work.\footnote{(8.2 seconds $\times$ 37,926 annotations); 3,600 seconds per hour; 8 hours per day $\to$ 10.8 days.} By employing an assistance system with only 50\% recall, the average time per annotation was reduced by about 1.7$\pm$3.6 seconds, which would have saved more than two work days in the illustrated example.\footnote{6.5 seconds per annotation $\to$ 8.6 days.}

In our evaluation we only examined three levels of recall for the automated assistance system, which already required a large number of study participants to get statistically meaningful results. In future work, it might be interesting to explore recall levels in between the studied boundaries to determine the sweet spot where a ML assistance model starts to facilitate the annotation workflow. Furthermore, it should be examined to what extent our results generalize to more complex tasks than NER, such as relationship extraction or POS tagging \cite{ringger2008assessing,fort2010influence}. Another important open question is how the different kinds of errors in the annotation process (Table~\ref{tab:annotationErrors}) influence the performance of the annotators. The performance of ML models is subject to a trade-off between precision and recall, therefore if certain errors in the pre-annotations are easier to correct than other (e.g.,\ it might be easier to remove a wrong annotation than to identify a missing one), the assistance system could be adapted with this in mind.

We hope that by improving the annotation process with our open source DALPHI framework, we can facilitate the creation of larger training datasets for more languages and specialized domains and therefore help bring (supervised) ML technologies to more people.

\section*{Acknowledgements}
  We would like to thank Dr. Nils Backhaus and the Chair of Cognitive Psychology and Cognitive Ergonomics of the Technische Universität Berlin for their support. DALPHI was developed by Hannes Korte, Arik Grahl, and Robert Greinacher, who were at the time employed by Implisense GmbH. Franziska Horn acknowledges funding from the Elsa-Neumann scholarship from the universities of Berlin.

\bibliography{dalphi_arxiv}
\bibliographystyle{acl_natbib}

\end{document}